\documentclass[preprint2]{aastex}

\newcommand{\snr}{{G7.5$-$1.7}}
\newcommand{\eg}{{3EG~J1809$-$2328}}
\newcommand{\axj}{AX~J1808.9$-$2310}
\newcommand{\asca}{{\it ASCA}}
\newcommand{\ros}{{\it ROSAT}}
\newcommand{\vla}{{\it VLA}}
\newcommand{\xmm}{{\it XMM}-Newton}
\newcommand{\kms}{km~s$^{-1}$}
\newcommand{\mjb}{mJy~beam$^{-1}$}

\newcommand{\HII}{H\,{\sc ii}}
\newcommand{\HI}{H\,{\sc i}}
\newcommand{\cc}{cm$^{-2}$}

\slugcomment{to appear in Astrophysical Journal}

\shorttitle{Discovery of a new X-ray Filled Supernova Remnant}
\shortauthors{Roberts and Brogan}

\begin{document}


\title{Discovery of a New X-ray Filled Radio Supernova Remnant Around the Pulsar Wind Nebula in 3EG J1809$-$2328}


\author{Mallory S.E. Roberts}
\affil{Eureka Scientific, Inc., Oakland, CA, USA}
\email{malloryr@gmail.com}

\and

\author{Crystal L. Brogan}
\affil{National Radio Astronomy Observatory, 520 Edgemont Rd,
  Charlottesville, VA 22903, USA}


\begin{abstract}

We report the discovery of a partial $\sim 2^{\circ}$ diameter non-thermal
radio shell coincident with Taz, the pulsar wind nebula (PWN) in the
error box of the apparently variable $\gamma$-ray source 3EG
J1809$-$2328. We propose that this radio shell is a newly identified
supernova remnant (SNR \snr) associated with the PWN. The SNR
surrounds an amorphous region of thermal X-rays detected in archival
\ros~ and \asca~ observations putting this system in the
mixed-morphology class of supernova remnants. \snr~ is the fifth such
supernova remnant coincident with a bright GeV source, and the fourth
containing a pulsar wind nebulae.
\end{abstract}




\section{Introduction}

The possibility of supernova remnants (SNRs) being sources of
observable GeV emission has long been suggested by theoretical models
of cosmic-ray acceleration \citep[see][and references therein]{ber+99}
in SNR shocks and by statistical associations with cataloged
$\gamma-$ray sources \citep{sd95,ehks96,yr97}. However, the direct linking of
a particular SNR with a specific $\gamma$-ray source detected by the
$EGRET$ instrument on board the Compton Gamma-Ray Observatory has been
elusive \citep{djg+03,trd+03} (note we only refer to the results of
the supernova blast as SNR, not isolated pulsar wind nebulae (PWN)
which are sometimes referred to as ``plerionic" or ``Crab-like" SNR).
The $EGRET$ sources with the smallest and most robust error boxes are
not positionally coincident with regions of SNRs which show direct
evidence of high-energy particle acceleration through synchrotron
emission of X-rays.  Although there are five previously known radio
SNRs that contain within their shells unidentified $EGRET$ sources
which are bright above 1 GeV \citep[as defined by][]{lm97}, in no case
can hard X-ray emission from the SNR blast wave be convincingly
associated with the $\gamma$-ray source \citep[][hereafter,
RRK]{rrk01}.

Interestingly, four of these SNRs are of the sub-class of
mixed-morphology SNR \citep[SNR][]{rp98} which have centrally
concentrated thermal X-ray emission within a radio shell.  Three of
these (CTA 1, IC 443, and W44) also contain young, energetic pulsars
which are actively producing high-energy particles as evidenced by
their wind nebulae \citep{brkc98,hgc+04,ocw+01,hhh96}.  It is
therefore unclear as to whether the GeV emission is produced at the
SNR forward shock, the interior where the thermal X-ray emission comes
from, or has to do with the pulsar or its wind.

While the source of GeV emission remains unclear, at TeV energies a
growing number of sources coincident with PWN and young shell SNR have
been discovered by H.E.S.S. \citep{aab+06}.  Several of the TeV
emitting PWN are associated with $EGRET$ sources, notably the Crab \citep{wcf+89},
Vela \citep{aab+06a}, and the two nebulae in the Kookaburra radio complex \citep{aab+06b}. Two clear
detections of shell SNRs with H.E.S.S. have also been reported
\citep{fun06}. In these cases, the TeV emission has been shown to be
associated with regions of non-thermal X-rays in the shell. A few
H.E.S.S. sources coincident with composite SNR have been reported as
well \citep{aaa+05,bgg+05,hgh+07}, but in these cases the TeV emission
appears to be associated with non-thermal X-rays from the central PWN,
not the shell.  Two of the mixed-morphology SNR associated with bright
GeV sources, IC443 and W28, have recently been detected as sources of
TeV emission, but it is unclear what, if any, connection there is
between the GeV source and the TeV source \citep{aaa+07,rbr+07}.

Here we report the detection of a non-thermal radio shell with the
Very Large Array (\vla\/) at 90~cm
surrounding the PWN associated with the $EGRET$ source GeV
J1809$-$2327/3EG J1809$-$2328. Archival \ros~ \citep{tru82} and \asca~ \citep{tih94} data show a
region of diffuse thermal X-ray emission between the radio shell and
the PWN suggesting this source is a new mixed-morphology SNR.

\section{Analysis of Archival X-ray Data in the Region surrounding 
the Taz PWN}

\begin{figure}
\epsscale{.98} \plotone{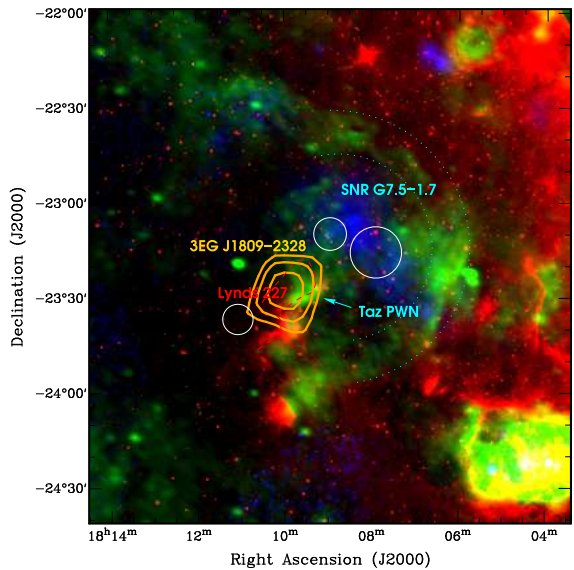}
\caption{\footnotesize{Multiwavelength view of \snr. MSX 8.3 $\mu$m ($20^{\prime \prime}$
  resolution) is in red, \ros~PSPC 0.1-2.4 keV X-rays 
 (smoothed with an $70^{\prime \prime}$ FWHM Gaussian) 
  are in blue, \vla\ 90~cm radio ($168^{\prime \prime} \times 120^{\prime \prime}$ resolution) is in
  green. The contours are the 68\%, 95\% and 99\% uncertainty on the
  position of \eg. The circles show X-ray spectral extraction regions (see fig. 2). The dotted cyan 
lines delineate the nominal shell of \snr~ while the dashed red line shows the approximate outline 
of the Lynds 227 dark cloud, whose edge is bright in the infrared image.  The 90~cm \vla\ image has been corrected for primary
  beam attenuation.} \label{multi}}
\end{figure}

3EG J1809$-$2328 is one of the brightest sources of $> 1$ GeV emission
in the Galaxy \citep{hbb+99,lm97}. It has a well determined position
(Fig~\ref{multi}) and apparently variable emission above 100 MeV on
timescales of months \citep{ntgm03}. \asca~ imaging of the $EGRET$
error box revealed an extended hard X-ray source (RRK), which
\citep{brrk02} subsequently resolved into an apparent rapidly-moving
PWN (RPWN) and a young stellar cluster using a short $Chandra$
observation \citep[see][for general reviews of PWN]{krh06,gs06}. Radio
imaging with the \vla\ and observations with \xmm\ confirm the
RPWN nature of the X-ray source \citep[][ Roberts et al. in
preparation]{rrk+01,brrk02}. Due to the radio nebula's distinctive
funnel shape (presumably imposed by its motion), its unusually
powerful $\gamma$-ray emission, and the growing tradition of naming
PWN after animals, this nebula is sometimes referred to as ``Taz"
(short for tasmanian devil).

\begin{figure}
\epsscale{.98} \plotone{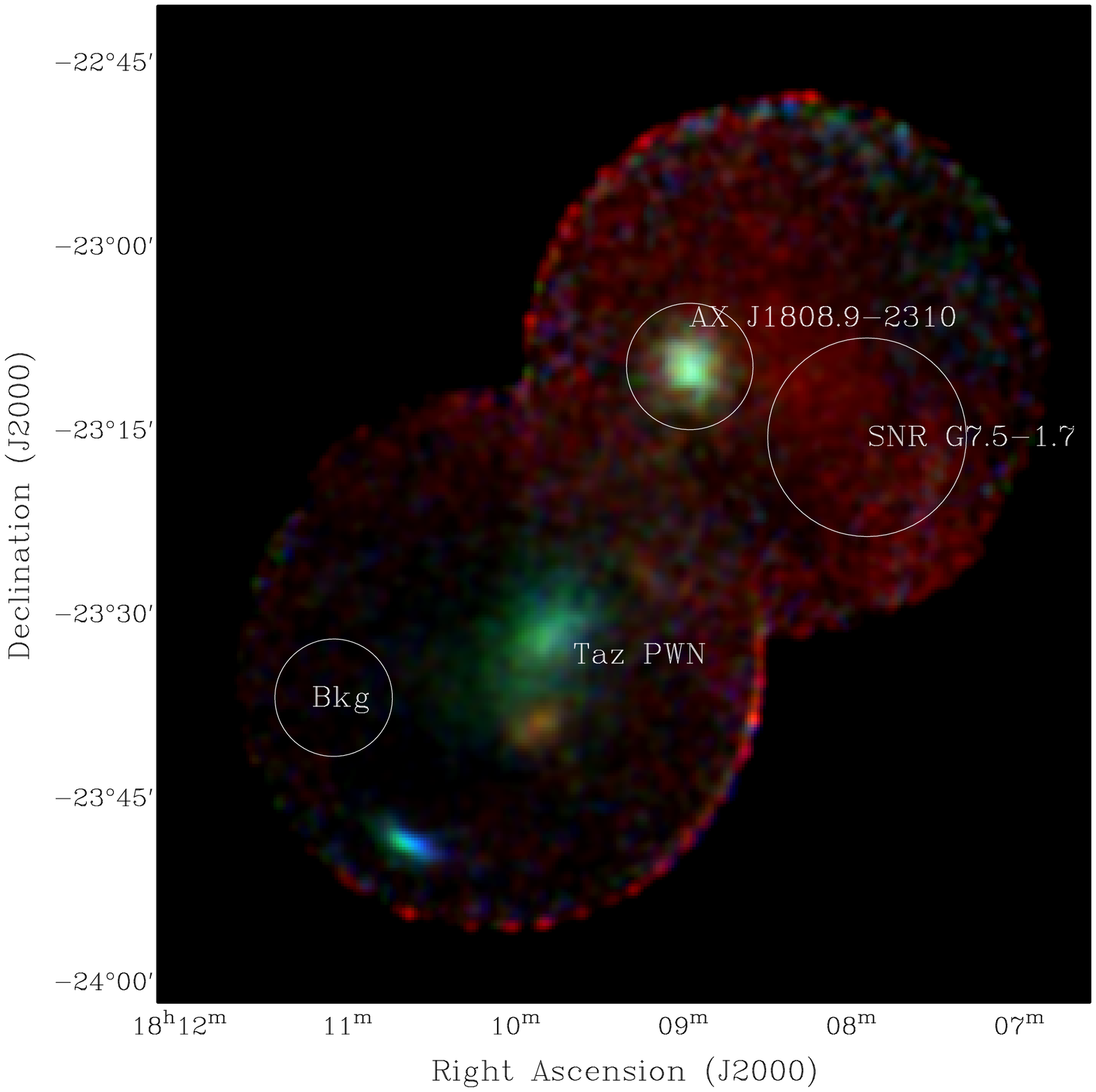}
\caption{\footnotesize{\asca~ GIS X-ray image of \snr. Red is 0.7-2~keV, green is
2-10~keV, blue is 4-10~keV ($\sim 1^{\prime}$ resolution). The circles show the extraction
regions for the X-ray spectral fits of the supernova remnant, the hard source, and the
background region.} \label{asca}}
\end{figure}

The X-ray and radio morphology of Taz, if interpreted as being created by a bowshock, 
suggests a birthsite to the
northwest.  A \ros~ PSPC exposure corrected 0.1-2.4 keV image obtained
through SkyView\footnote{The NASA SkyView Virtual Observatory is
located at http://skyview.gsfc.nasa.gov/} shows a large region of soft
X-ray emission in this direction which we will refer to as \snr~
(Fig.~\ref{multi}). This inspired a 38~ks observation with the \asca~
satellite on 1996-09-24 that has not previously been
published. Following the imaging and spectral fitting procedures
outlined in RRK, we produced combined particle background subtracted,
exposure corrected images of the \eg~ and \snr~ fields from the \asca~
GIS data in the 0.7-2 keV, 2-10 keV, and 4-10 keV bands
(Fig.~\ref{asca}). The higher energy bands show a fairly bright,
compact source of hard X-rays in the region of soft thermal X-ray
emission which is not evident in the \ros~ image. The peak of this
source is at RA = $18^h 08^m 56.3^s$, Dec = $-23^{\circ} 09^{\prime}
57^{\prime \prime}$ (J2000) in the GIS 2-10 keV image; \asca\ images have a
nominal positional error of $\sim 24^{\prime \prime}$ \citep{guf+00}.
We will refer to this hard source as \axj.

We extracted spectra from two regions in the field of the GIS
instruments, one containing the hard source and one containing the
soft diffuse emission. Since there is emission essentially throughout
the \asca~ field, we used a background region from the neighboring
field which contains the Taz PWN. We fit the soft extended
emission to a thermal model, choosing an absorbed $vnei$
non-equilibrium model from $XSPEC$ to facilitate comparison with other
Mixed-Morphology remnants, at first keeping the abundances fixed to solar.  
To help constrain the soft end of the
thermal emission spectrum, we simultaneously fit the \ros~ PSPC
spectrum extracted from the the same region. We obtained an adequate
fit using a single temperature with solar abundances. The result is 
a temperature $kT=0.61^{+0.07}_{-0.08}$
keV (90\% confidence region), an ionization timescale
$\tau=1.54^{+0.80}_{-0.47}\times 10^{11}\,{\rm s\, cm^{-3}}$ and an
absorption $n_H = 1.5^{+0.8}_{-0.6}\times 10^{21}$~\cc\/. Note that the \ros~ and \asca~
spectra seemed to differ somewhat near the upper end of the \ros~ passband ($ \sim 2$ keV) and
so these values are somewhat dependent upon which energy bins from \ros~ were included in
the fits. For example, ignoring the  \ros~ data above 1.6 keV resulted in an improved chi-squared
and a somewhat lower absorption of $ \sim 9\times 10^{20}$~\cc\/. Fitting the
\asca~ data by themselves resulted in consistent values, and allowing the
abundances to vary did not result in a significantly better fit. 
We also tried the $ray$ and $vray$ equilibrium thermal
models, but these resulted in statistically inferior fits. The spectral 
values obtained are typical of temperatures and timescales fit from \asca~
observations of other mixed-morphology SNR \citep{kon+05}.  Since the
thermal X-ray emission covers an area much larger than the field of
view of the \asca~ GIS and it is somewhat unclear what is the total
extent even in the \ros~ images, an accurate total thermal flux
measurement is difficult. However, using the fluxed \ros~ maps and
extrapolating from our fit spectra, we estimate the total unabsorbed bolometric thermal flux
from the remnant to be on the order of $10^{-10} {\rm erg\, cm^{-2}\,
s^{-1}}$.

We then fit the hard source (\axj) to an absorbed power law plus an
absorbed $vnei$ model, fixing the latter to the values above allowing
only the normalization to vary. The result is an absorption $n_H =
1.9^{+0.5}_{-0.4}\times 10^{22}$~\cc\/, a spectral index $\Gamma=
2.57^{+0.29}_{-0.27}$ and a 2-10 keV flux $F_X=3.9\pm 0.1 \times 10^{-12}~{\rm ergs \,
cm^{-2}\, s^{-1}}$. Since there is no evidence for a source of
additional foreground absorption in the mid-infrared or optical (see
below), the order of magnitude higher absorption toward this source
suggests it is an unrelated background object and/or contains significant intrinsic
absorption. The total Galactic $n_H$ in this direction 
as estimated using the $HEASARC$ $n_H$ tool and the \HI\/ maps of \citet{dl90} is $\sim 
6\times 10^{21}$~\cc\/. 

\section{Radio Observations}

We observed the region around Taz at 90~cm with the \vla\ on May 28,
2004 in the DnC configuration (project code AR 547) for 4.5 hours
(on-source).  We used 32 channels over a bandwidth of 12.6 MHz in LL
and RR polarizations.  Using AIPS and standard wide-field low
frequency imaging techniques \citep[see for example][and references
therein]{bdl+04}, the resulting image has an effective central
observing frequency of 324.84~MHz, an effective bandwidth of 9.38~MHz,
a beam size of $168^{\prime \prime}\times 120^{\prime \prime}$
(P.A.=$72\arcdeg$) and an RMS level of about 20~mJy/beam. The \vla\
FWHP primary beam at 325 GHz is $2.5\arcdeg$. Due to spatial filtering
caused by the missing short spacings, this image is not sensitive to
smooth structures larger than $\sim 1\arcdeg$.

\begin{figure}
\epsscale{.98} \plotone{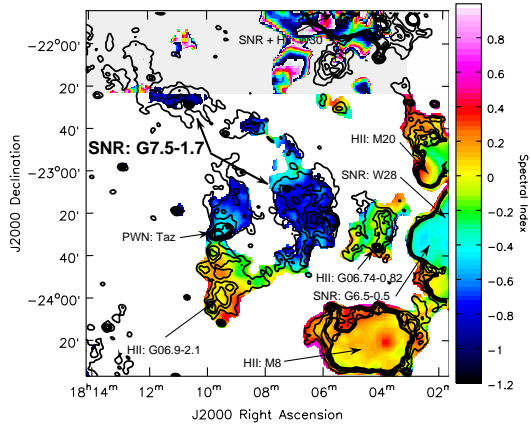}
\caption{\footnotesize{Spectral index (assuming $S_{\nu}^{\alpha}$) map of \snr~ and
the surrounding region derived from an 11~cm Bonn image and the 90~cm
\vla\ image. Before calculating the spectral index, the Bonn image was
passed through a ``high-pass" filter to mimic the spatial coverage of
the \vla\ interferometric data (which are not sensitive to smooth
strucures larger than $1\arcdeg$). The 90~cm image was first convolved
to the $258''$ resolution of the Bonn image. The input 11 and 90cm
images were masked at 30 and 120 \mjb\/, respectively. Contours from
the $168''\times 120''$ resolution 90~cm image are superposed, the
contour levels are $60\times$(1, 2, 3, 4, 5) \mjb\/. Several
previously identified \HII\/ regions and SNRs are labeled for
reference \citep{gwbe94,lph96,bgg+06}.}}
\label{spimage}
\end{figure}

The 90~cm image shows a partial shell of emission surrounding the
thermal X-ray emission (Fig.~\ref{multi}). Comparison with the $8.3\mu$
MSX image of the region shows no infrared excess associated with this
shell, suggesting a non-thermal origin of the radio emission
\citep{cg01,bgg+06}.  We convolved the 90~cm image with a 2-D
Gaussian to produce a $258^{\prime \prime}$ resolution image for
comparison to the Bonn 11cm Survey image for this region
\citep{rfh+84}. We made two spectral index maps from
the 90~cm and 11~cm images (assuming $S_\nu^\alpha$), one using the
standard 11~cm background subtracted image, and one using an 11~cm
image that was passed through a ``high-pass" filter in order to mimic
the large-scale spatial filtering of the 90~cm \vla\ image. In both
spectral index maps, the Taz PWN and the new shell both have
significantly steeper spectra than known thermal sources in the
region. Overall, the spectral index map created from the filtered 11cm
Bonn data (Fig.~\ref{spimage}) is steeper, as expected, and provides a
better match to the spectra of other known SNR in the field of view.
The spectral index of the shell ranges from $\alpha \sim -0.3$\ to
$-0.7$, with an estimated uncertainty of $\pm 0.3$ at any particular
position. The large uncertainty is dominated by the inherent
uncertainty in measuring flux from a large, diffuse object with an
interferometer, and not the noise in the images.

The center of the radio shell is at roughly $l=7.54^{\circ},\, b=-1.90^{\circ}$ near
the northern edge of the Taz radio nebula, and the radius is $r \sim
0.82^{\circ}$. The shell has a somewhat smaller radius of curvature
($r\sim 0.45^{\circ}$ center at $l\sim 7.51, \, b \sim -1.42$) on the
side nearer the Galactic plane, centered in the middle of the thermal
X-ray emission. Assuming the origin of the expanding shell lies
somewhere in between these two centers and near the symmetry axis of
Taz, we will refer to the radio shell as \snr.

We have also obtained high resolution ($3.6^{\prime \prime} \times
3.2^{\prime \prime}$) 20~cm \vla\ data of a much smaller field of view
centered on the Taz nebula which will be discussed in detail in a
future paper that concentrates on the PWN (Roberts et al. in
prep.). For the purpose of the present paper it is interesting to note
that there is a 13 mJy 20~cm radio source at RA(J2000) = 18$^{\rm h}$08$^{\rm
  m}$57.59$^{\rm s}$, Dec(J2000) = $-23^{\circ} 09^{\prime} 45.4^{\prime
  \prime}$ that is coincident with the hard \asca~ source \axj.  The
radio source is elongated and well fit by a $7.4^{\prime \prime}
\times 1.4^{\prime \prime}$ Gaussian, suggestive of a barely resolved
double jet source.

\section{Discussion}

\snr~ has the classic properties of a mixed-morphology supernova
remnant \citep{rp98}. The thermal X-ray emission is concentrated
between Taz and the Galactic plane side of the radio shell.  In both
the Bonn 11~cm and our \vla\ 90~cm images, between 1/2 and
3/4 of a radio shell surrounds Taz. The side towards the Galactic
plane is brightest, while the side away from the plane is too faint to
be sure of any structures, suggesting a strong density gradient in
this direction.  Comparing to the MSX $8.3\mu$m image, the radio shell
is located within a region of low diffuse mid-infrared emission.
Numerical models of the evolution of mixed-morphology remnants suggest
they are a common stage in the growth of remnants at ages of
$\sim$10,000-100,000~yr if they are expanding in high-density regions
of the ISM \citep{tbh06}. In this case where there is an apparent density 
gradient, the thermal X-ray emission is concentrated near the bright side of the shell, 
towards the Galactic plane, rather than around the apparent center of the shell. 

The position of the center of \snr~ suggests it is the birth remnant
of Taz.  While the overall X-ray absorption of Taz is significantly
higher ($n_H \sim 1.8 \times 10^{22} \, {\rm cm}^{-2}$, RRK) it is also
moving into or behind the Lynds 227 dark cloud. \xmm\ imaging
shows some parts of the X-ray nebula are absorbed more than others
\citep{rgh+06}, and a preliminary spectral analysis suggests a range
of $n_H \sim 0.7-2.0 \times 10^{22} \, {\rm cm}^{-2}$.  If we assume
Taz and \snr\ are associated, then the distance to Lynds 227 of
1.7~kpc \citep{okn+99} can be considered a lower limit to \snr. 

From the above estimates of the center of the radio shell, and the
pulsar wind nebula line of symmetry, we estimate the birthsite of Taz
to be $l\sim 7.53^{\circ}, b\sim -1.68^{\circ}$. The average transverse space velocity
is then implied to be $v_T\sim 200(d_{1.7}/t_{50})$~\kms\/, fairly
typical for an isolated pulsar. Here $d_{1.7}$ is the distance in
units of 1.7~kpc and $t_{50}$ is the age in units of 50~kyr. 
The distance from this estimated birthsite to the pulsar
is about half the average distance to the shell ($R_b\sim 0.62^{\circ}$), 
which is roughly where a bowshock is expected to begin
forming around a pulsar moving within a decelerating SNR shell \citep{vdk04}.  

If we accept the association with the pulsar and hence the distance to Lynds 227 as a lower 
limit on the SNR distance, and  we further assume it has been undergoing
Sedov expansion for the majority of its existence (i.e. $R\propto
t^{0.4}$) then we can estimate the current expansion velocity of the shell to be $v_s
\sim 150(d_{1.7}/t_{50})$~\kms\/. Note that if the 
SNR were considerably younger (eg. $t \la 15,000$~yr), or considerably further away ($d \ga 5$~kpc), 
then the velocity of the radio shell would be similar to very young SNRs. For example, 
only a few degrees away on the sky is the X-ray and radio bright shell of the $\sim 2000$yr old 
Type II remnant SNR G11.2$-$0.3 which has a well measured expansion 
velocity of $\sim 700$~\kms\/ at a distance of $\sim 5$kpc \citep{rtk+03,tr03}. 
For strong shocks, the temperature behind the shock, which in the case of G11.2$-$0.3 is
$kT \sim 0.6$~keV, is proportional to the square of the shock
velocity. Although the flux from such a shock depends somewhat on the density and composition of the ISM as well as the equilibrium state of the electrons with the ions, such a hot shock would be 
difficult to hide from ROSAT given the rather bright radio emission and overall low absorption. 
However, at the estimated velocity of $\sim 150$km/s given our nominal distance and age, the bulk of 
the shock emission would be emitted well below 1 keV where absorption is more important and 
ROSAT is much less sensitive. Therefore, the lack of bright X-ray emission associated 
with the radio shell is consistent with the interpretation of \snr~ being at a distance of 
$\sim 2$~kpc and an age similar to other mixed-morphology SNR. 

\snr~ is now the fifth known mixed-morphology supernova remnant that
is coincident with the error box of one of the 25 brightest sources of
GeV emission in our Galaxy, and the fourth of these to contain a
pulsar wind nebula. Two of these SNRs also contain known TeV sources,
and several other PWN associated with bright GeV sources have TeV
emission \citep{frth07}, so it is quite plausible that Taz/\snr\ also has TeV
emission. However, in no currently known case is the GeV emission
convincingly arising from the same spatial location as the TeV
emission. This suggests that GeV and TeV emission may have separate
sites of acceleration or represent particle populations with very
different ages.

As is the case with the PWN in CTA1 and W44, Taz is actually contained
within the GeV error box, while the radio shells are outside the error
box. In all three of these cases, there is some evidence of moderate
variability of the $\gamma$-ray emission, with the source containing
Taz showing the strongest evidence \citep{ntgm03}. The W44/PSR
B1853+01 system is very similar in that there is clearly an RPWN seen
in both radio and X-rays at the edge of the thermal X-ray emitting
region. It also has similar X-ray \citep{kon+05} and $\gamma$-ray
properties (RRK).  The rapid motion of the pulsars through the clumpy,
dense interior of these remnants provides a natural variability
mechanism for emission coming from the immediate environment of the pulsar. 
However, PSR B1853+01 is only moderately energetic with a
spin-down power $\dot E = 4.3\times 10^{35}$~erg~s$^{-1}$. Therefore, if this is where 
the variable $\gamma$-rays from W44 are produced, then it suggests
the interaction of a pulsar wind with the unique hot and dense
environment of the interior of mixed-morphology remnants may allow for
very efficient $\gamma$-ray production.
 
\acknowledgments

This research has made use of data obtained from the High Energy Astrophysics Science Archive Research Center (HEASARC), provided by NASA's Goddard Space Flight Center. 
The National Radio Astronomy Observatory Very Large Array is a facility of the National Science Foundation operated under cooperative agreement by Associated Universities, Inc. 
This research made use of data products from the Midcourse Space 
Experiment,  processing of which data was funded by the Ballistic 
Missile Defense Organization with additional support from NASA 
Office of Space Science.  This research has also made use of the 
NASA/ IPAC Infrared Science Archive, which is operated by the 
Jet Propulsion Laboratory, California Institute of Technology, 
under contract with the National Aeronautics and Space 
Administration.



{\it Facilities:} \facility{MSX}, \facility{Effelsberg}, \facility{ASCA (GIS)}, \facility{VLA}, \facility{ROSAT (PSPC)}

\bibliographystyle{apj}
\bibliography{journals_apj,myrefs,psrrefs,crossrefs,egret}

\end{document}